\documentclass[11pt]{article}
\usepackage[margin=1in]{geometry}
\usepackage[T1]{fontenc}
\usepackage{lmodern}
\usepackage{amsmath,amssymb}
\usepackage{upgreek}
\usepackage{graphicx}
\usepackage{booktabs}
\usepackage{xcolor}
\usepackage[numbers,sort&compress]{natbib}
\usepackage[colorlinks=true,linkcolor=blue!50!black,citecolor=blue!50!black,urlcolor=blue!50!black]{hyperref}
\usepackage[section]{placeins}
\usepackage{flafter}
\usepackage{caption}
\graphicspath{{figures/}}

\newcommand{\um}{\,\ensuremath{\upmu}m}

\newcommand{\ThBareShort}{13.4}               %
\newcommand{\ThBareMid}{7.7}                  %
\newcommand{\ThBareLong}{2.5}                 %
\newcommand{\ThSlope}{0.061}                  %
\newcommand{\NgWG}{1.75}                      %
\newcommand{\BeamWidth}{9.2}                  %
\newcommand{\SpreadK}{0.44}                   %
\newcommand{\SpreadAir}{6.0}                  
\newcommand{\SpreadOx}{4.2}                   %
\newcommand{\DelayFs}{51}                     %
\newcommand{\TBPsi}{18}                       %
\newcommand{\Qeven}{\ensuremath{27\mbox{--}45}}  %
\newcommand{\Qodd}{\ensuremath{16\mbox{--}24}}
\newcommand{\FIBone}{4.25}                     %
\newcommand{\BWdev}{176}                      %
\newcommand{\BWdevBare}{95}                   %
\newcommand{\FIBdevBare}{8.25}                %
\newcommand{\CEBWdev}{123.1}                  
\newcommand{\CEBWdevBare}{66.3}              

\newcommand{\FIBdev}{4.25}                     %

\newcommand{\PKdev}{\ensuremath{-1.55}}
\newcommand{\PKdevBare}{\ensuremath{-1.57}}
\newcommand{\WalkLimDev}{88}                  %
\newcommand{\RatioDev}{1.85}                   %
\newcommand{\SpreadDevBare}{10.6}
\newcommand{\SpreadDevOne}{3.8}
\newcommand{\InsMean}{0.22}                   %
\newcommand{\YorderFull}{0.06}                
\newcommand{\BWfdy}{120}                       
\newcommand{\BWfdyBare}{77}                    
\newcommand{\PKfdy}{\ensuremath{-1.11}}        
\newcommand{\PKfdyBare}{\ensuremath{-1.08}}    
\newcommand{\RatioFdy}{1.56}                   
\newcommand{\CEBWfdy}{93.0}                   

\title{Ultra-broadband Huygens metasurface-integrated grating coupler}
\author{\parbox{0.95\textwidth}{\centering
  Junhe~Zhang\textsuperscript{1,\textdagger}, Roy~Avrahamy\textsuperscript{1,\textdagger}, Hanyu~Zheng\textsuperscript{2}, Cosmin-Constantin~Popescu\textsuperscript{2}, Fan~Yang\textsuperscript{2}, Xu~Fang\textsuperscript{3}, Tian~Gu\textsuperscript{2}, and Juejun~Hu\textsuperscript{1,2}\\[4pt]
  {\small \textsuperscript{1}Department of Materials Science and Engineering, Massachusetts Institute of Technology, 77 Massachusetts Avenue, Cambridge, MA 02139, USA\\[2pt]
  \textsuperscript{2}2Pi Inc., Cambridge, MA 02139, USA\\[2pt]
  \textsuperscript{3}School of Electronics and Computer Science, University of Southampton, Southampton~SO17~1BJ,~UK\\[2pt]
  \textsuperscript{\textdagger}These authors contributed equally to this work.}}}

\date{}

\begin{document}
\maketitle

\begin{abstract} 
\noindent Grating couplers provide a common optical interface for photonic integrated circuits because they are compatible with foundry fabrication, wafer-level testing, and scalable high-density packaging. 
However, their narrow operating bandwidth, due to the grating's inherent wavelength-dependent diffraction response, is their Achilles' heel. 
In this paper, we propose and design a scheme to significantly expand the bandwidth of grating couplers by adding a Huygens metasurface layer on top to correct the angular dispersion of diffracted beams. 
We present detailed device operation mechanisms and a physics-informed rational design methodology for constructing such a design. 
Full-wave simulations show that our design nearly doubles the bandwidth of a state-of-the-art SiN/Si dual-layer grating coupler without using a metal reflector. 
The design achieves a minimum coupling loss of \PKdev~dB and a 1-dB bandwidth of \BWdev~nm covering the entire S, C, and L telecom bands. 
We further perform a detailed fabrication tolerance analysis showing that the design's tolerance is well within the capabilities of standard Si foundry fabrication. 
\end{abstract}

\section{Introduction}
\label{sec:intro}

\begin{figure}[!tbp]
  \centering
  \includegraphics[width=0.506\linewidth]{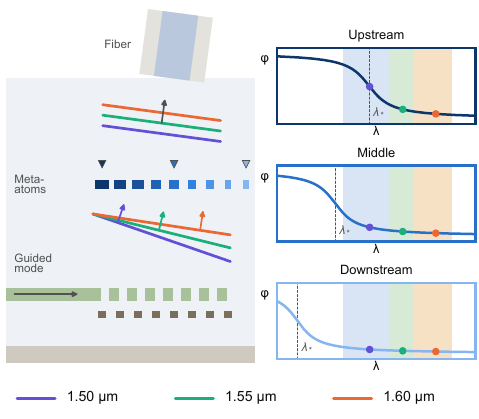}
  \caption{Schematic of the metasurface-integrated (henceforth referred to as ``MS-integrated'') grating coupler. Each wavelength leaves the grating at its own angle. The metasurface contains an array of resonant Huygens meta-atoms whose resonance wavelengths $\lambda_*$ are graded along the grating. Above the metasurface, the phase fronts of all wavelengths are nearly parallel and enter the fiber at a similar angle. The insets show the transmission phase $\phi$ of the upstream, middle, and downstream meta-atoms, marked by triangles of the same color. Downstream, the same phase curve moves to shorter $\lambda_*$. Dots mark the three wavelengths of the legend. In the insets, shading marks the S, C, and L bands.}
  \label{fig:problem}
\end{figure}

Grating couplers are a standard way to move light between an optical fiber and a photonic chip: they need no polished facet,
they can sit anywhere on the wafer, and they allow testing before dicing \cite{marchetti2019prj,ranno2022acsphotonics}. Their
weakness is bandwidth. A grating couples the guided mode to free space through the phase-matching condition
$\sin\theta = n_\mathrm{eff}(\lambda) - \lambda/\Lambda$, where $\theta$ is the air-equivalent emission angle, $\lambda$ is the free-space wavelength, $n_\mathrm{eff}$ is the effective index of the guided mode, and $\Lambda$ is the grating period. Each wavelength therefore leaves at a different angle. Because of this angular walk-off \cite{xiao2013oe}, a fiber held at one angle collects only part of the spectrum (Fig.~\ref{fig:problem}).

One route to a wider band reduces the angular dispersion itself through lowering the group index of the guided mode, for example using silicon
nitride and subwavelength-engineered waveguides \cite{cheben2023aop}. Along this line, dual-layer SiN/Si couplers have emerged as a particularly promising platform given their compatibility with standard Si photonic foundry fabrication \cite{sacher2014oe,mak2018oe,xu2017ol,Fraser2026}. By pairing a low-group-index nitride waveguide with a reflective silicon high-contrast grating beneath it to enhance directionality, Sacher \textit{et al.} reported a SiN-on-SOI coupler reaching $-1.3$~dB minimum coupling loss with an 80-nm 1-dB bandwidth in the C band
\cite{sacher2014oe}, and a recent demonstration by TSMC attained $-1.16$~dB loss and a 62-nm 1-dB bandwidth in the
O band \cite{tsmc_ofc2026}. In this study, we base our design on this type of grating optimized for both low loss and large bandwidth following foundry design rules. The grating coupler's geometric parameters are summarized in the Supplementary Information Section S1. Full-wave 3D simulations of its spectral response give an optimized 1-dB bandwidth of \BWdevBare~nm. We refer to this grating coupler without the metasurface as the ``baseline'' design.

A second route compensates the dispersion with an element placed after the grating, for instance a prism or a second grating \cite{spaulding1991ao,klenk2020,zhu2026cjl}. We note that while several proposals have combined metasurfaces with grating couplers, they serve different purposes, such as beam deflection and focusing \cite{wu2023photonics,xie2024ol,qu2023oe}, rather than dispersion compensation. This gap can be explained by the relatively small dispersion in classical optical metasurfaces, where the phase delay is accumulated through light propagation in waveguide-type meta-atoms. Compensating for the angular dispersion of typical grating couplers would require deep subwavelength waveguide-type meta-atoms well more than 5\um{} thick (Supplementary Sec.~S2), which is simply unrealistic for practical fabrication.

Here we propose, for the first time to our knowledge, to use resonant Huygens metasurfaces to enable dispersion compensation in grating couplers. Unlike waveguide meta-atoms, Huygens metasurfaces capitalize on two spectrally overlapping quasi-normal modes with opposite longitudinal symmetries and matching linewidths to achieve broadband near-unity transmission and a $2\pi$ phase excursion across the resonance---a phenomenon known as the Kerker effect \cite{kerker1983,pfeiffer2013,decker2015}. This strong wavelength dependence of phase can also be realized within a deeply subwavelength thickness~\cite{zhang2018huygens,fathnan2020huygens,zhang2020ultrathin,shalaginov2021metalens}, making Huygens metasurfaces ideally suited for dispersion compensation in grating couplers.

Implementing the design, however, must address two challenges. First, each meta-atom in the metasurface must simultaneously meet two requirements: control its local transmission phase $\phi$ \emph{relative to neighboring meta-atoms} to match the tilted wavefront of the fiber, and provide the phase dispersion needed for angular compensation, characterized by the group delay $\tau=\partial\phi/\partial\omega$, where $\omega$ is the angular frequency. While a library of meta-atoms can readily be selected to meet either requirement alone, as in typical metasurface applications, meeting both concurrently mandates careful dispersion engineering that also accounts for fabrication compatibility and tolerance. Second, changes in the incidence angle can detune the two resonances that overlap at normal incidence, severely degrading the transmission efficiency of a Huygens metasurface \cite{arslan2017jpd}. For a standard single-mode fiber with a mode-field diameter of 10.4\um{} at 1550~nm \cite{corning2025smf28}, the Gaussian-mode angular spectrum extends to approximately $\pm5.4^\circ$ in air-equivalent angle at the $1/e^2$ intensity level. This angular spread is sufficiently large that the angle-dependent degradation must be properly mitigated.

In the following, we elaborate the device physics and design methodology for addressing the two challenges described above (Section 2). We then implement the design in a dual-layer SiN/Si grating coupler operating near 1550~nm (Section 3). Full-wave simulations show that the Huygens layer nearly doubles the coupler's 1-dB bandwidth to \BWdev~nm while achieving a minimum coupling loss of \PKdev~dB. Fabrication and alignment tolerances are examined in the Supplementary Information.

\section{Device physics and design methodology}
\label{sec:concept}

\subsection{Angular walk-off and the delay gradient that cancels it}
We take $x$ along the grating, $y$ across it, and $z$ normal to the chip. With the time convention $e^{-i\omega t}$, where $t$ denotes time, the grating emits light with in-plane wavevector component ${k_x(\omega) = \beta(\omega) - 2\pi/\Lambda}$,
where $\beta$ is the propagation constant of the guided mode. We denote the $x$ component of the wavevector accepted by a fiber tilted at the air-equivalent angle $\theta_f$ by $k_x^{f}(\omega) = (\omega/c)\sin\theta_f$, where $c$ is the speed of light. For this air-equivalent angle, $\sin\theta_f$ is conserved across the index-matched layers. Matching the two at the center frequency $\omega_0$ leaves a wavevector mismatch $\Delta k(\omega)$ that grows
linearly with detuning \cite{xiao2013oe},
\begin{equation}
  \Delta k(\omega) = k_x^{f}(\omega) - k_x(\omega) \simeq \frac{\omega-\omega_0}{c}\,\left(\sin\theta_f - n_g\right),
  \label{eq:mismatch}
\end{equation}
where $n_g = c\, d\beta/d\omega$ is the group index of the grating's Bloch mode. On the coupler studied here the beam propagates at
\ThBareShort$^\circ$ at 1.46\um{} and \ThBareLong$^\circ$ at 1.63\um, about $-\ThSlope^\circ$/nm. These air-equivalent angles are computed from the 2D full-wave field of the bare grating. Differentiating $\sin\theta = n_\mathrm{eff} - \lambda/\Lambda$ gives
$d\sin\theta/d\lambda = -(n_g - \sin\theta)/\lambda$, so this slope, with $\theta \approx \ThBareMid^\circ$ at 1.55\um, corresponds to $n_g \approx \NgWG$. A Gaussian fiber mode of
$1/e^2$ radius $w_f$, fed by a matched Gaussian beam, tolerates a wavevector mismatch of $\pm0.96/w_f$ for a 1-dB loss. With
Eq.~\eqref{eq:mismatch}, this tolerance gives an estimate for the 1-dB bandwidth $\Delta\lambda_\mathrm{1dB}$,
\begin{equation}
  \Delta\lambda_\mathrm{1dB} \approx \frac{0.96\,\lambda^2}{\pi\,w_f\,(n_g-\sin\theta_f)} \approx \WalkLimDev~\mathrm{nm}
  \label{eq:walkofflimit}
\end{equation}
for $\lambda = 1.55$\um{} and the baseline's fiber angle $\theta_f = \FIBdevBare^\circ$. Here $w_f = 5.2$\um{} is half the 10.4-\ensuremath{\upmu}m mode-field diameter of SMF-28. Our full-wave simulation yields a 1-dB bandwidth of \BWdevBare~nm for the
baseline grating coupler (Section~\ref{sec:results}), close to this estimate. A lower group index slows the growth of the mismatch and so widens the band. For comparison, a grating coupler to a silicon waveguide, with $n_g \approx 4.2$, would spread the beam over about 28$^\circ$ from 1.45 to 1.63\um, against 11$^\circ$ here. Equation~\eqref{eq:walkofflimit} then gives a 1-dB bandwidth of about 35~nm, against \WalkLimDev~nm for the SiN waveguide grating coupler here.

Next we consider adding a thin layer on the grating that adds a wavelength-dependent wavevector to compensate the wavevector mismatch. A layer with transmission phase $\phi(x,\omega)$ adds $\partial\phi/\partial x$ to the in-plane wavevector. We write
$\phi(x,\omega) = \phi_0(x) + (\omega-\omega_0)\,\tau(x)$, where $\phi_0(x)$ is the transmission phase at $\omega_0$ and $\tau(x) = \partial\phi/\partial\omega$ at $\omega_0$ is the group delay that the metasurface imposes at position $x$. The added
wavevector is then $\partial\phi_0/\partial x + (\omega-\omega_0)\,\partial\tau/\partial x$. The
wavelength-independent part $\partial\phi_0/\partial x$ deflects the beam and so changes the optimal fiber angle. This part changes the walk-off only through the term $\sin\theta_f$ of Eq.~\eqref{eq:mismatch}, which is small here (about 0.07 for a \FIBdev$^\circ$ fiber angle, as compared to $n_g = \NgWG$). Its frequency-dependent part cancels the mismatch of Eq.~\eqref{eq:mismatch} when
\begin{equation}
  c\,\frac{\partial\tau}{\partial x} = \sin\theta_f - n_g ,
  \label{eq:delaygradient}
\end{equation}
where the right-hand side is dimensionless. In our case, Eq.~\eqref{eq:delaygradient} specifies that the delay must fall along the propagation direction by about 5.6~fs per
micron, or \DelayFs~fs across the \BeamWidth-\ensuremath{\upmu}m beam ($1/e^2$ intensity width at the metasurface, set by the grating and narrower than the 10.4-\ensuremath{\upmu}m fiber mode). Equivalently, the angularly dispersed beam of the grating carries a tilted pulse front \cite{hebling1996oqe}, and the metasurface
must impose the opposite pulse-front tilt.

\subsection{Huygens metasurface that supplies the delay}
\label{sec:huygens}
A thin layer without resonances delays light only by its optical thickness \cite{presutti2020optica}: \DelayFs~fs would need
several micrometers of silicon (Supplementary Sec.~S2): we therefore opt for resonant metastructures.

\begin{figure}[!tp]
  \centering
  \includegraphics[width=\linewidth]{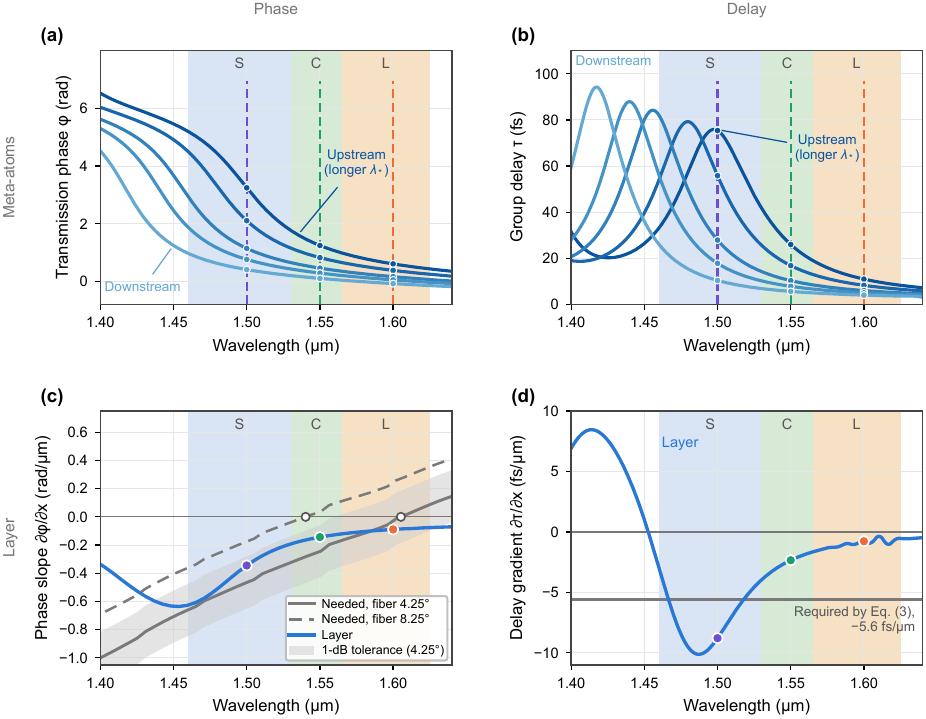}
  \caption{Phase (left) and group delay (right) of single meta-atoms (top) and of the metasurface layer (bottom), computed from the angular-spectrum-averaged transmission (Eq.~\eqref{eq:teff}). (a) Transmission phase, with a common offset for all meta-atoms, and (b) group delay of five example meta-atoms, with $\lambda_*$~=~1.420, 1.442, 1.460, 1.482, and 1.499\um. Darker curves belong to upstream meta-atoms with longer $\lambda_*$. Dashed lines and dots mark 1.50, 1.55, and 1.60\um{} in all panels. (c) Phase gradient supplied by the metasurface (blue) and required at the air-equivalent fiber angle of \FIBone$^\circ$ (gray) and at the baseline fiber angle of \FIBdevBare$^\circ$ (dashed). The gray band is the fiber's 1-dB tolerance, $\pm0.96/w_f$. Open circles mark where the required curves cross zero. (d) Delay gradient supplied by the metasurface (blue) and the requirement of Eq.~\eqref{eq:delaygradient} at \FIBone$^\circ$ (gray). The gradients in (c) and (d) are linear fits over the 24 central sites, weighted by beam power. Shading marks the S, C, and L bands.}
  \label{fig:concept}
\end{figure}

For a Huygens meta-atom supporting two overlapping resonances, the ideal complex transmission amplitude $t(\omega)$ and its group delay $\tau(\omega)$ are \cite{fan2003josaa,suh2004apl}
\begin{equation}
  \begin{aligned}
    t(\omega) &= \exp\!\left[2i\arctan\!\frac{2Q(\omega-\omega_*)}{\omega_*}\right],\\
    \tau(\omega) &= \frac{\partial\arg t(\omega)}{\partial\omega}
    = \frac{4Q/\omega_*}{1+\left[2Q(\omega-\omega_*)/\omega_*\right]^2}.
  \end{aligned}
  \label{eq:allpass}
\end{equation}
Here $Q$ is the resonance quality factor, $\omega_* = 2\pi c/\lambda_*$ is the resonance frequency, and $\lambda_*$ is the resonance wavelength. The peak delay is $4Q/\omega_*$, whereas the full width at half maximum of the delay response is $\omega_*/Q$. Increasing $Q$ therefore raises the peak delay while narrowing its bandwidth. The integrated delay, $\int_{-\infty}^{\infty}\tau(\omega)\,d\omega=2\pi$, is independent of $Q$. For an approximately constant delay difference $\Delta\tau$ across a band of angular-frequency width $\Delta\omega$, one resonance pair per meta-atom thus has an ideal delay--bandwidth limit of $|\Delta\tau|\Delta\omega\le 2\pi$. Figure~\ref{fig:concept}b shows the group delay for five meta-atoms of the layout, each with the $2\pi$ phase budget.

However, we note that our design can compensate for the grating's angular dispersion over a bandwidth considerably larger than would be expected from the $2\pi$ delay--bandwidth limit, since exact cancellation of angular walk-off is not required for coupling into a fiber of finite angular acceptance. The fiber's 1-dB wavevector acceptance, $\pm0.96/w_f$, permits a phase difference of about $\pm1.7$~rad across the \BeamWidth-\ensuremath{\upmu}m beam. In our case, for the \DelayFs-fs delay difference required across the same beam (Eq.~\eqref{eq:delaygradient}), the ideal $2\pi$ phase budget falls short of exact compensation over the \BWdev-nm 1-dB bandwidth by about 0.9~rad, within what the fiber acceptance can accommodate.

For the same beam, fiber angle, and \BWdev-nm band, a grating coupler to a silicon waveguide, with $n_g\approx4.2$, would require $|\Delta\tau|\Delta\omega\approx\TBPsi$~rad (Eq.~\eqref{eq:delaygradient}). This is nearly three times the $2\pi$ phase budget of one pair, and the remaining mismatch would greatly exceed the fiber's acceptance. Under these conditions, multiple resonance pairs would be needed. The low group index of the SiN waveguide mode is therefore essential to the single-layer design here.

\subsection{Concurrent dispersion and phase control with one metasurface layer}
\label{sec:dispphase}
\label{sec:principles}

Equation~\eqref{eq:delaygradient} sets the dispersion requirement: the metasurface must supply a group-delay gradient that decreases along the grating. We prioritize this requirement by grading the resonance wavelength $\lambda_*$ of the Huygens meta-atoms. On the long-wavelength side of their resonances, meta-atoms resonant at longer $\lambda_*$ have larger delay (Fig.~\ref{fig:concept}b); a downstream decrease in $\lambda_*$ therefore produces a delay gradient of the required sign (Fig.~\ref{fig:concept}b,d).

Once the resonance gradient is chosen, the phase requirement must be met with the same meta-atoms. With one resonance pair per meta-atom, phase and group delay follow the same resonant response (Eq.~\eqref{eq:allpass}), so selecting $\lambda_*$ to set the delay also fixes the relative phase between meta-atoms. For meta-atoms of similar $Q$, phase curves shifted in resonance wavelength do not cross (Fig.~\ref{fig:concept}a); their separation decreases on the long-wavelength side of the resonances. The graded metasurface thus imparts a phase slope of one sign that fades toward long wavelengths (Fig.~\ref{fig:concept}c, blue). Ideal coupling into a fiber at angle $\theta_f$ requires this slope to equal $k_x^f-k_x$. At the baseline fiber angle of \FIBdevBare$^\circ$ (Fig.~\ref{fig:concept}c, dashed), however, the required slope changes sign at the center of the band, near 1.54\um{}. The resonance grading that supplies the delay gradient cannot also match this sign-changing phase slope.

\begin{figure}[!t]
  \centering
  \includegraphics[width=\linewidth]{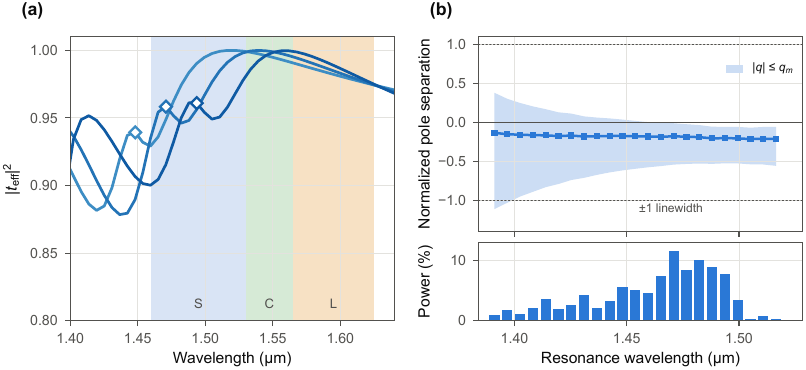}
  \caption{(a) Transmission $|t_\mathrm{eff}|^2$ of three example meta-atoms ($\lambda_*$~=~1.448, 1.471, and 1.494\um), averaged over the beam's angular spectrum (Eq.~\eqref{eq:teff}). Open diamonds mark each meta-atom's design resonance wavelength $\lambda_*$. Shading marks the S, C, and L bands. (b) The signed difference between the even- and odd-resonance wavelengths of the meta-atoms, normalized by the even-resonance linewidth $\lambda_*/Q_\mathrm{e}$. Squares give this difference at the grating coupler's central emission angle ($q = 0$). The band shows its range over $|q| \le q_m$, where $q_m = \SpreadK$~rad/\ensuremath{\upmu}m is the $1/e^2$ half-width of the beam's angular spread. The bottom panel gives each meta-atom's share of the beam power incident on the layout's 40 sites.}
  \label{fig:layer}
\end{figure}

We therefore select the fiber angle to bring the required phase slope into alignment with that supplied by the graded metasurface. Turning the fiber from $\theta_f$ to $\theta_f'$ changes $k_x^f$ by $(\omega/c)(\sin\theta_f'-\sin\theta_f)$, shifting the required slope by an approximately wavelength-independent amount. At \FIBone$^\circ$, its zero moves to about 1.61\um{}, near the long-wavelength edge where the metasurface's phase slope has faded (Fig.~\ref{fig:concept}c). This choice only weakly changes the target delay gradient, since $n_g$ dominates $\sin\theta_f$ in Eq.~\eqref{eq:delaygradient}. The metasurface's phase slope then remains within the fiber's 1-dB acceptance around the required slope from about 1.44 to 1.63\um{}. With a free fiber angle, one metasurface layer thereby meets the dispersion and phase requirements simultaneously.

For a predefined fiber angle, a second, non-resonant metasurface layer or a baseline grating optimized for a larger angle gives the same constant offset (Supplementary Secs.~S7 and~S8).

\subsection{Angular tolerance of the Huygens metasurface}
\label{sec:angular}
The beam leaving the grating and incident upon the metasurface is \BeamWidth\um{} wide ($1/e^2$ intensity), and its angular spectrum has a $1/e^2$
half-width of $\pm\SpreadOx^\circ$ in the
oxide ($\pm\SpreadAir^\circ$ air-equivalent) at 1.504\um. Every meta-atom thus sees a range of incidence angles. In a Huygens metasurface, the Huygens condition requires the even and odd poles to coincide, and tilting the incident light moves them at different rates. In previously reported silicon-nanodisk Huygens metasurfaces, the electric and magnetic dipole resonances separate away from normal incidence, resulting in transmission dips
\cite{arslan2017jpd}. Let $q$ denote the deviation of the in-plane wavevector from the value corresponding to the grating coupler's central emission angle at each wavelength. To first order in $q$, the even and odd resonance wavelengths vary as $\lambda_\mathrm{e,o}(q) \approx \lambda_* + v_\mathrm{e,o}\,q$, where $v_\mathrm{e,o}$ are their respective slopes with respect to $q$. The poles then separate by $|v_\mathrm{e} - v_\mathrm{o}|\,|q|$. A meta-atom then stays close to all-pass only when this separation is small compared with the linewidths of the two resonances. The even resonance has the higher quality factor $Q_\mathrm{e}$, so its linewidth $\lambda_*/Q_\mathrm{e}$ sets the limit.

To address this angular sensitivity, we systematically investigated how to engineer $v_\mathrm{e}-v_\mathrm{o}$ (Supplementary Sec.~S3). The principal finding is that this difference is governed mainly by the metasurface lattice ratio $Y/(2a)$, which compares the row spacing $Y/2$ along $y$ with the pitch $a$ along $x$ (Fig.~\ref{fig:device}b). Minimizing $|v_\mathrm{e}-v_\mathrm{o}|$ alone, however, increases the linewidth mismatch between the two Huygens resonances, departing from the Kerker condition and increasing reflection loss. Therefore, we choose a $Y/(2a)$ ratio that balances linewidth matching against the resonance detuning induced by angular dispersion. Coupled with the broad resonances of the Huygens meta-atoms ($Q_\mathrm{e}$~=~\Qeven{} and $Q_\mathrm{o}$~=~\Qodd), this metasurface lattice choice keeps the resonance separation within one even-resonance linewidth for most meta-atoms (Fig.~\ref{fig:layer}b). Appreciable separation occurs only for meta-atoms with smaller $\lambda_*$, which receive a small fraction of the incident beam power (Fig.~\ref{fig:layer}b, bottom panel). The finite angular spread of the incident beam therefore incurs little transmission loss (Fig.~\ref{fig:layer}a).
\subsection{Design method}
\label{sec:method}
Our design methodology, informed by the physical insights discussed in the preceding sections, follows the workflow in Fig.~\ref{fig:device}c. A single simulation of the grating coupler without the metasurface provides the input field on the metasurface plane. The same simulation gives the beam's central emission angle and angular spread at each wavelength. The meta-atom library is computed at these angles and averaged over this spread. A model based on the local periodic approximation (LPA) \cite{pestourie2018oe,perezarancibia2018oe} then combines the field and the library to choose the metasurface layout. During this search, 3D periodic simulations, which cost far less than the full device, calibrate the model (Supplementary Sec.~S5). A 3D full-wave simulation of the complete focusing device finally checks the design (Section~\ref{sec:device}). The grating coupler and the metasurface can be designed separately because the grating's evanescent field has decayed over the 2.28-\ensuremath{\upmu}m gap.

\textit{Input field.} A 2D finite-difference time-domain (FDTD) simulation (Tidy3D \cite{tidy3d,hughes2021apl}) launches the
guided mode and records the $y$ component of the electric field, $E_y$, and the $x$ component of the magnetic field, $H_x$, on the plane of the metasurface at 45 wavelengths from 1.39 to 1.64\um. The up-going part of this field,
normalized to the guided power, is the input of everything that follows.

\textit{Meta-atom library.} At each wavelength, each meta-atom is solved by rigorous coupled-wave analysis (RCWA) \cite{moharam1995josaa} (fmmax \cite{fmmax}) at the grating coupler's central emission angle and at one angle on each side of it, which sample the beam's angular spread (Eq.~\eqref{eq:teff}). Huygens meta-atoms are designed by pinning both complex poles \cite{lalanne2018lpr} of the
even and odd resonances at the target wavelength $\lambda_*$ \cite{chen2026oe,zhang2026arxiv} (Supplementary Sec.~S4). The library has 25 super-ellipses with $\lambda_*$ from 1.380 to 1.517\um, 5.7~nm apart, and 8 non-resonant disks. Each
meta-atom is tabulated at 45 wavelengths. Its effective transmission $t_\mathrm{eff}$ is the complex transmission averaged over a Gaussian model of the beam's angular spectrum with a three-point Gauss--Hermite rule, \begin{equation} t_\mathrm{eff}(\lambda) = \tfrac{2}{3}\,t(k_x) + \tfrac{1}{6}\left[t(k_x+q_1) + t(k_x-q_1)\right],\qquad q_1=\sqrt{3}\,q_m/2, \label{eq:teff} \end{equation} where $q_m = \SpreadK$~rad/\ensuremath{\upmu}m is the measured $1/e^2$ half-width of the beam's angular spectrum at the metasurface (Section~\ref{sec:angular}).

\textit{Meta-atom selection.} The LPA treats each site of the metasurface as part of an infinite array of identical copies \cite{pestourie2018oe,perezarancibia2018oe}. The incident field is multiplied inside each site by that meta-atom's
$t_\mathrm{eff}$, propagated through the oxide with the angular spectrum method, and overlapped with the fiber mode at the fiber
plane to obtain the coupling efficiency $\eta$:
\begin{equation}
  \eta(\lambda) = \kappa\,\frac{\left|\sum_k A(k)\,G(k)\,\psi_f^*(k)\,\mathrm{Re}\,k_z\right|^2}{\sum_k |\psi_f(k)|^2\,\mathrm{Re}\,k_z},
  \label{eq:overlap}
\end{equation}
where $k$ indexes the in-plane wavevectors, $A$ is the angular spectrum after the metasurface, $G$ the propagation to the fiber plane in the index-matched region above the
oxide, $\psi_f$ the fiber mode, $k_z$ the vertical wavevector, and $\kappa$ the normalization to the guided power.

Within the LPA, the field coupled into the fiber can be decomposed into contributions from individual sites, so the effect of replacing the meta-atom at one site is computed exactly and cheaply. The layout search maximizes a score that rewards high mean coupling and a 1-dB band covering the S, C, and L bands. It starts from the library layout closest to the best layout of ideal Huygens meta-atoms. Each site in turn takes the library meta-atom that most improves the score, as in direct search optimization \cite{yang2021ol,yang2023nanoph}. Limiting the resonance-wavelength difference between neighboring meta-atoms keeps the LPA accurate and suppresses diffraction into non-zero orders (Supplementary Sec.~S6). The model is calibrated against 3D periodic runs of the layout until the two agree (Supplementary Sec.~S5).

\section{Broadband low-loss metasurface-assisted grating coupler design}
\label{sec:implementation}

\subsection{Device and simulation setup}
\label{sec:device}
\begin{figure}[!t]
  \centering
  \includegraphics[width=\linewidth]{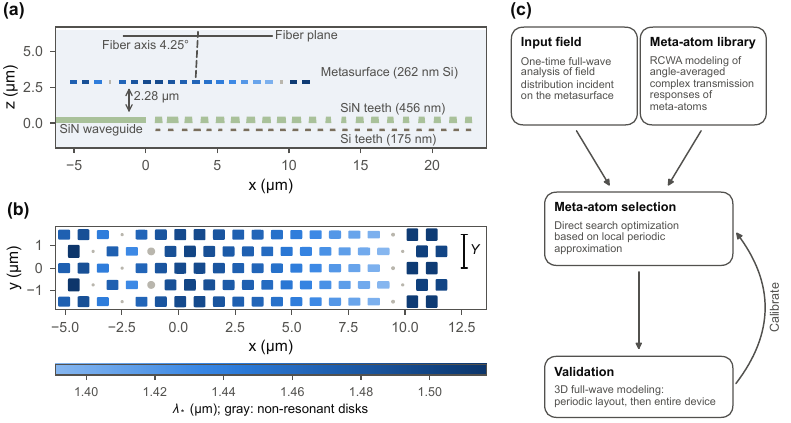}
  \caption{(a) Cross-section of the MS-integrated grating coupler through the meta-atom row at $y = 0$, drawn to scale with the
  geometry of the 3D simulations. The fiber axis at which the MS-integrated device is read out (\FIBdev$^\circ$, air-equivalent) is drawn at its angle in the oxide. (b) Top view of the metasurface over two periods in $y$. The meta-atoms form a centered rectangular lattice with pitch $a = 0.854$\um{} along $x$ and two offset rows per $y$-period $Y$, marked by the bracket. The colors give each meta-atom's resonance wavelength
  $\lambda_*$, and gray marks the non-resonant disks. (c) Design workflow.}
  \label{fig:device}
\end{figure}

The baseline coupler is a dual-layer SiN/Si grating on a 2.28-\ensuremath{\upmu}m buried oxide (Fig.~\ref{fig:device}a). Its 456-nm SiN and 175-nm Si layers are separated by 398~nm of oxide, and the grating pattern is etched through the full thickness of both layers (Supplementary Table~S1). The grating teeth are modeled with 85$^\circ$ sidewall angles to represent realistic etch profiles. The grating is fed by the TE mode of the SiN waveguide. The top of the oxide is index-matched to the fiber (epoxy attach \cite{snyder2013tcpmt}), so there is no oxide--air reflection.

The metasurface is a single 262-nm-thick layer of silicon meta-atoms, embedded in the oxide with its bottom 2.28\um{} above the top of the SiN. The resonant meta-atoms have super-elliptical outlines close to rounded rectangles, and the meta-atoms are arranged on a centered rectangular lattice with pitch $a = 0.854$\um{} along $x$ and period $Y = 1.479$\um{} along $y$. Each period contains two rows separated by $Y/2$ along $y$ and offset by $a/2$ along $x$. For identical meta-atoms, this lattice keeps every non-zero diffraction order
evanescent in the oxide
(Supplementary Sec.~S3). Forty sites, from $x = -5.0$ to 11.6\um{} ($x = 0$ marks the start of the grating), cover the beam emitted by the grating with margin (Fig.~\ref{fig:device}b). The 24 central sites receive 95\,\% of the incident power and hold resonant Huygens meta-atoms. Their $\lambda_*$ rises over the first four sites, where the beam builds up, and then ramps down from 1.499 to 1.391\um{} along the propagation direction. The 16 sites at the two ends receive the remaining 5\,\% and also contain five non-resonant meta-atoms shaped as disks. An alternative design using only resonant meta-atoms reaches nearly the same level of performance (Supplementary Sec.~S6).

\begin{figure}[!tbp]
  \centering
  \includegraphics[width=0.99\linewidth]{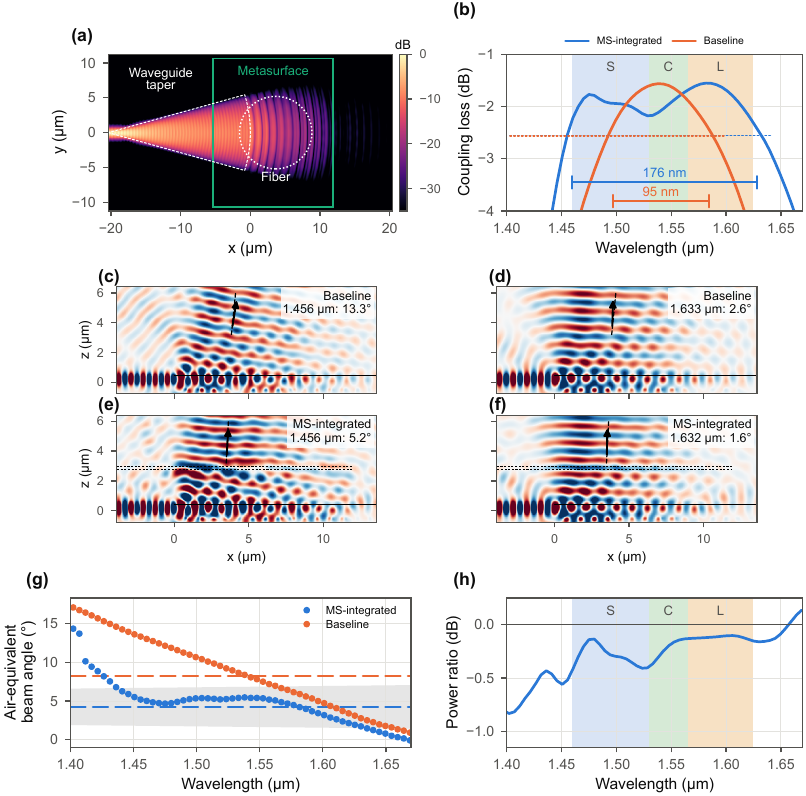}
  \caption{(a) Simulated $|E_y|^2$ (in dB) in the SiN plane of the full MS-integrated device at 1.550\um. Dashed lines outline the taper, which ends at the start of the grating ($x = 0$), and the dotted circle marks the fiber mode ($1/e^2$ radius). (b) Simulated coupling loss of the MS-integrated (blue) and baseline (orange) couplers. Brackets mark the full devices' 1-dB bands, and dotted lines lie 1~dB below each device's minimum coupling loss. (c--f) Real part of the electric field's $y$ component in the $x$--$z$ plane at $y=Y/4$, between meta-atom rows, at about 1.456 and 1.632\um, near the band edges, for the baseline (c, d) and MS-integrated (e, f) devices. The metasurface is outlined. Arrows and dashed lines show the beam direction and fiber axis in oxide; labels give air-equivalent angles. (g) Air-equivalent beam angle. Dashed lines in the devices' colors mark the fiber angles giving the widest 1-dB bands for the MS-integrated and baseline devices; gray band marks the former's 1-dB angular acceptance. (h) Upward power through the fiber plane, in MS-integrated coupler relative to the baseline device. In (b) and (h), shading marks the S, C, and L bands.}
  \label{fig:fulldevice}
\end{figure}

We simulate the focusing grating and overlying metasurface in three dimensions using finite-difference time-domain calculations (Tidy3D 2.12). The guided TE mode enters through a tapered SiN waveguide, and the grating teeth follow iso-phase arcs \cite{vanlaere2007ptl} (Fig.~\ref{fig:fulldevice}a; Supplementary Sec.~S5). Coupling efficiency is calculated from the overlap of the field at the fiber plane with a tilted Gaussian representation of the SMF-28 mode, normalized to the launched guided power. We simulate the baseline grating without the metasurface and calculate its fiber coupling in the same way. All simulations, including the RCWA of the meta-atoms, use constant refractive indices taken at 1.55\um{}: 1.444 for SiO$_2$ \cite{malitson1965}, 3.475 for Si \cite{li1980si}, and 1.998 for LPCVD SiN \cite{luke2015}. Repeating the 3D periodic simulation of the final layout with dispersive indices gives a nearly identical 1-dB bandwidth (Supplementary Sec.~S5).

\subsection{Coupler Performance}
\label{sec:results}

The full device simulation (Fig.~\ref{fig:fulldevice}a) shows that the metasurface widens the 1-dB bandwidth \RatioDev-fold, from \BWdevBare~nm for the baseline to \BWdev~nm for the MS-integrated device, with a minimum coupling loss of \PKdev~dB against \PKdevBare~dB (Fig.~\ref{fig:fulldevice}b and Table~\ref{tab:summary}). For each device, we select the fiber angle and position that give the widest 1-dB band without noticeably lowering the peak coupling, and with margin against dips that could split the band (Supplementary Sec.~S5).

Fig.~\ref{fig:fulldevice}c--f explains why the MS-integrated device extends the bandwidth. At 1.456\um{}, the metasurface bends the beam from the baseline device's 13.3$^\circ$ to 5.2$^\circ$, whereas at 1.632\um{} it changes the angle only from 2.7$^\circ$ to 1.6$^\circ$. The stronger steering at shorter wavelengths narrows the range of beam angles across the band from \SpreadDevBare$^\circ$ to \SpreadDevOne$^\circ$, which fits within the fiber's 1-dB angular acceptance of about $\pm2.6^\circ$ (Fig.~\ref{fig:fulldevice}g). At the short-wavelength edge, however, the meta-atoms approach resonance and the resulting loss brings the coupling below the 1-dB threshold before angular walk-off becomes limiting (Fig.~\ref{fig:fulldevice}h). At the long-wavelength edge, the metasurface provides too little steering to maintain alignment with the fiber, and angular mismatch sets the bandwidth limit.

The metasurface has an average insertion loss of about 0.3~dB over its 1-dB band. About \InsMean~dB of upward power is lost, mostly to reflection caused by the unequal linewidths of each meta-atom's two resonances (Fig.~\ref{fig:fulldevice}h). Another \YorderFull~dB is diffracted into the $\pm1$ orders along $y$, which the fiber does not collect, because the two offset rows hold slightly different meta-atoms. Correcting the angular dispersion more than recovers this loss near the band edges.

\begin{table}[!htb]
  \centering
  \caption{Simulated performance of the baseline and MS-integrated couplers. The 1-dB bandwidth is measured 1~dB below the spectrum's own peak.}
  \label{tab:summary}
  \footnotesize\setlength{\tabcolsep}{3pt}
  \begin{tabular}{@{}lccccc@{}}
    \toprule
    & \begin{tabular}[b]{@{}c@{}}Fiber angle\\($^\circ$)\end{tabular}
    & \begin{tabular}[b]{@{}c@{}}1-dB band\\nm (edges, \ensuremath{\upmu}m)\end{tabular}
    & \begin{tabular}[b]{@{}c@{}}Minimum coupling\\loss (dB)\end{tabular}
    & \begin{tabular}[b]{@{}c@{}}Band-averaged\\loss (dB)\end{tabular}
    & \begin{tabular}[b]{@{}c@{}}Worst loss across\\S+C+L (dB)\end{tabular} \\
    \midrule
    Baseline      & 8.25 & \begin{tabular}[c]{@{}c@{}}95\\(1.493--1.588)\end{tabular}  & $-1.57$ & $-1.90$ & $-4.61$ \\
    MS-integrated & 4.25 & \begin{tabular}[c]{@{}c@{}}176\\(1.456--1.632)\end{tabular} & $-1.55$ & $-1.92$ & $-2.32$ \\
    \bottomrule
  \end{tabular}
\end{table}

\section{Discussion and conclusion}
\label{sec:discconc}

\subsection{Discussion}
\label{sec:discussion}
\begin{figure}[!t]
  \centering
  \includegraphics[width=\linewidth]{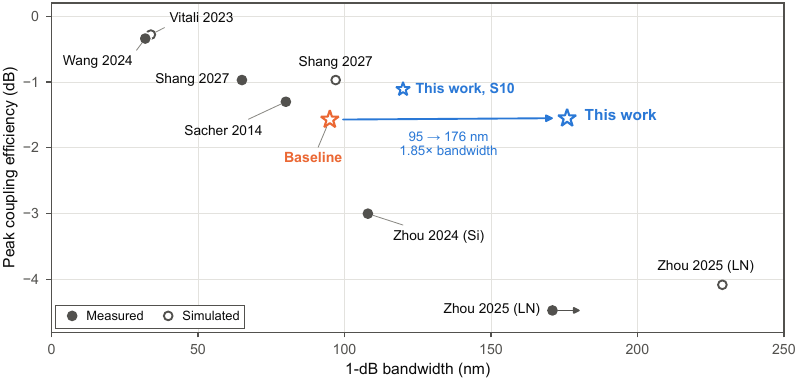}
  \caption{Peak coupling efficiency versus 1-dB bandwidth for selected grating couplers near 1550~nm. Filled circles denote measurements, open circles denote simulations, and stars denote our results. The smaller star marks an alternative design discussed in Supplementary Sec.~S10. Short arrows mark bandwidths reported as lower bounds. The plotted values are listed in Table~\ref{tab:benchmark}, and each study's fiber, polarization, and evaluation method in Supplementary Table~S5. The additional shorter-wavelength entries are tabulated but not plotted.}
  \label{fig:literature-benchmark}
\end{figure}

\begin{table}[!t]
  \centering
  \caption{Representative high-performance grating couplers that couple directly to a standard single-mode fiber. Simulated entries are full 3D simulations. The coupling efficiency (CE)--bandwidth (BW) product uses the linear peak coupling efficiency and $\Delta\lambda_{1550}$. Entries near 1550~nm are shown at their reported bandwidths; shorter-wavelength entries are scaled from their respective operating wavelengths $\Delta\lambda_{1550}=\Delta\lambda_{1\mathrm{dB}}(1550/\lambda_0)$. Further comparison conditions are given in Supplementary Table~S5.}
  \label{tab:benchmark}
  \scriptsize\setlength{\tabcolsep}{2pt}
  \begin{tabular}{@{}p{0.30\linewidth}ccccc@{}}
    \toprule
    Coupler / variant & Result & Peak CE (dB) & BW reported (nm) & BW at 1550 (nm) & CE $\times$ BW at 1550 (nm) \\
    \midrule
    Wang et al.\ \cite{wang2024sciadv} & Meas. & $-0.34$ & $32$ & $32$ & $29.6$ \\
    Shang et al.\ \cite{shang2026optcom} & Meas. & $-0.97$ & $65$ & $65$ & $52.0$ \\
    Shi et al., O band\ \cite{tsmc_ofc2026} & Meas. & $-1.16$ & $62$ & $73.4$ & $56.2$ \\
    Sacher et al.\ \cite{sacher2014oe} & Meas. & $-1.30$ & $80$ & $80$ & $59.3$ \\
    Mak et al., O band\ \cite{mak2018oe} & Meas. & $-2.20$ & $72.9$ & $86.3$ & $52.0$ \\
    Zhou and Tsang\ \cite{zhou2024broadband} & Meas. & $-3.00$ & $108$ & $108$ & $54.1$ \\
    Zhou et al.\ \cite{zhou2025nanophotonics} & Meas. & $-4.47$ & $\geq171$ & $\geq171$ & $\geq61.1$ \\
    \midrule
    Vitali et al.\ \cite{vitali2023prj} & Sim. & $-0.28$ & $33.9$ & $33.9$ & $31.8$ \\
    Shang et al.\ \cite{shang2026optcom} & Sim. & $-0.97$ & $97$ & $97$ & $77.6$ \\
    Zhou et al.\ \cite{zhou2025nanophotonics} & Sim. & $-4.08$ & $229$ & $229$ & $89.5$ \\
    \midrule
    This work, baseline & Sim. & $\PKdevBare$ & $\BWdevBare$ & $\BWdevBare$ & $\CEBWdevBare$ \\
    This work, MS-integrated & Sim. & $\PKdev$ & $\BWdev$ & $\BWdev$ & $\CEBWdev$ \\
    This work, S10 & Sim. & $\PKfdy$ & $\BWfdy$ & $\BWfdy$ & $\CEBWfdy$ \\
    \bottomrule
  \end{tabular}
\end{table}

To benchmark the MS-integrated coupler's performance against existing designs, we use the coupling efficiency--bandwidth product $P_{\mathrm{CE-BW}}=\eta_{\mathrm{pk}}\Delta\lambda_{1\mathrm{dB}}$, where $\eta_{\mathrm{pk}}$ is the peak coupling efficiency in linear power units and $\Delta\lambda_{1\mathrm{dB}}$ is the 1-dB bandwidth. Adding the metasurface increases this product from \CEBWdevBare{} to \CEBWdev~nm. Figure~\ref{fig:literature-benchmark} and Table~\ref{tab:benchmark} compare our results with published grating couplers that couple directly to a standard single-mode fiber and were measured or simulated in full 3D (Supplementary Sec.~S11). As shown in Fig.~\ref{fig:literature-benchmark}, the baseline device is already among the best of these couplers in the coupling efficiency--bandwidth trade-off, and the metasurface-integrated device extends the Pareto front further: to the best of our knowledge, its simulated \CEBWdev-nm efficiency--bandwidth product is the highest of any coupler in this comparison.

The proposed device geometry is compatible with CMOS foundry fabrication: it builds on a foundry-compatible dual-layer SiN/Si grating and adds a single patterned Si metasurface embedded in oxide. A detailed fabrication tolerance study (Supplementary Sec.~S9) shows that the S, C, and L bands remain within the 1-dB bandwidth for meta-atom thicknesses of 257--272~nm and tested grating-to-metasurface gaps of 0.75--5.00\um. This band coverage is also retained for metasurface overlay errors of $\pm100$~nm and fiber position errors within 1.25\um{} in any direction. Meta-atom size and sidewall slope change impose tighter constraints, although systematic size bias can be calibrated and a known sidewall slope can be incorporated into the meta-atom design.

Further improvements can target the two principal sources of metasurface insertion loss identified in Sec.~\ref{sec:results}: reflection caused by unequal linewidths of the overlapping resonances and diffraction into nonzero transverse orders arising from the different meta-atoms in the two offset rows. A meta-atom library with better-matched resonance linewidths, together with an optimization that accounts explicitly for reflection and the two-row supercell response, may reduce these losses while preserving the required group-delay and phase profiles. Moreover, the metasurface is designed here for a predefined grating coupler structure, showing that our design scheme can directly adapt to existing grating couplers without modifying them. As another example, on a grating coupler from a different foundry stack, the same procedure widens the 1-dB bandwidth \RatioFdy-fold, from \BWfdyBare~nm to \BWfdy~nm, at nearly the same peak coupling efficiency (\PKfdy~dB against \PKfdyBare~dB; Supplementary Sec.~S10). Co-design of the grating coupler and metasurface will likely improve the performance further.

\subsection{Conclusion}
\label{sec:conclusion}

To the best of our knowledge, this work is the first to combine a metasurface with a grating coupler to enhance its coupling bandwidth through angular-dispersion compensation. We present a physics-informed, rational design methodology that establishes the required group-delay gradient, selects Huygens meta-atoms to provide the corresponding dispersion, and matches the phase profile by optimizing the fiber angle. The design flow uses the grating coupler's simulated output field and can be readily adapted to different coupler geometries, operating wavebands, and process design kit configurations.

In full-device three-dimensional simulations, the metasurface-integrated SiN/Si coupler achieves a \BWdev-nm 1-dB bandwidth covering the S, C, and L bands, with a peak coupling efficiency of \PKdev~dB. Its \CEBWdev-nm coupling efficiency--bandwidth product is, to the best of our knowledge, the highest among grating couplers measured or simulated in full 3D with a standard single-mode fiber. The device adds only one patterned Si metasurface to a foundry-compatible grating stack and is amenable to CMOS foundry fabrication. The design retains S/C/L coverage across the evaluated variations in meta-atom thickness, grating-to-metasurface spacing, overlay, and fiber alignment. Such dispersion-engineered metasurface-integrated grating couplers could make broadband, efficient fiber packaging a practical capability across a wide range of foundry-fabricated photonic circuits.

\section*{Data availability}
Data underlying the results may be obtained from the authors upon reasonable request.

\bibliographystyle{unsrtnat}
\bibliography{refs}

\end{document}